\documentclass[10pt,english,a4paper,showpacs,amsmath,amssymb,floatfix,twocolumn]{revtex4-1}
\usepackage{amsmath,amstext,amssymb,graphicx,babel,geometry,setspace,bm,dcolumn,color}
\usepackage[colorlinks=true, linkcolor=red, citecolor=blue, urlcolor=blue, 
linktoc=page, bookmarks=true, pdfstartview={FitH}, pdfborder={0 0 0.0 [3 3]}]{hyperref}
\usepackage[T1]{fontenc}
\begin{document}
\title{Noise-induced Regime Shifts: A Quantitative Characterization}
\author{Sayantari Ghosh}
\email{sayantari@jcbose.ac.in}
\author{Amit Kumar Pal} 
\email{amit@jcbose.ac.in}
\author{Indrani Bose}
\email{indrani@jcbose.ac.in}
\affiliation{Department of Physics, Bose Institute, 93/1, A. P. C. Road, Kolkata-700009, India}

\begin{abstract}
Diverse complex dynamical systems are known to exhibit abrupt regime shifts at bifurcation points 
of the saddle-node type. The dynamics of most of these systems, however, have a stochastic 
component resulting in noise driven regime shifts even if the system is away from the 
bifurcation points. In this paper, we propose a new quantitative measure, namely, the 
propensity transition point as an indicator of stochastic regime shifts. The concepts and the 
methodology are illustrated for the one-variable May model, a well-known model in ecology and the genetic toggle, a two-variable 
model of a simple genetic circuit. The general 
applicability and usefulness of the method for the analysis of regime shifts is further demonstrated in the case of the
mycobacterial switch to persistence for which experimental data are available.  
\end{abstract}


\maketitle

\section{Introduction}
\label{intro}

\begin{figure}
 \begin{center}
  \includegraphics[scale=0.7]{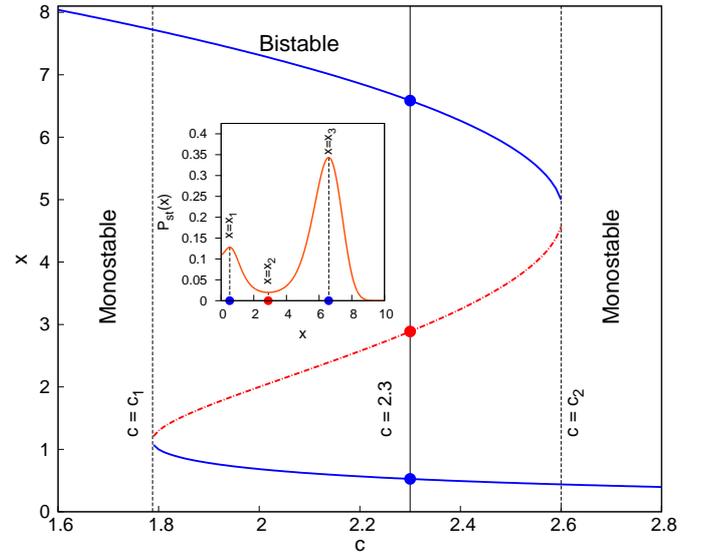}
 \end{center}
\caption{\small{Biomass $x$ versus the maximum grazing rate parameter $c$ in the 
steady state of the May model the dynamics of which are described by Eq. (\ref{model}). 
The solid lines represent the stable steady states and the dot-dashed line the branch 
of unstable steady states. The points $c=c_{1}$ and $c=c_{2}$ represent the lower 
and upper bifurcation points respectively. The parameter values used are $r=1$, 
$K=10$ and $x_{0}=1$. The inset shows the plot of the steady state probability 
distribution $P_{st}(x)$ versus $x$ in the presence of additive noise of strength 
$d_{1}=.25$. The maximum grazing rate parameter $c=2.3$. The extrema points 
correspond to the stable steady states of the deterministic dynamics, with $x_{1}$ 
and $x_{3}$ representing the stable steady states which are separated by an unstable 
steady state $x_{2}$.}}
\label{fig1}
\end{figure}

Dynamical systems are known to undergo sudden regime shifts at critical parameter values, termed the 
bifurcation points, with the different regimes defined by distinct sets of attractors of the dynamics \cite{strogatz94}. 
Recently, a large number of studies have been devoted to the investigation of regime shifts in complex dynamical 
systems and processes ranging from ecosystems, climate change, population collapse and  financial markets 
\cite{scheffer091,scheffer092,scheffer03,lenton11}
to epileptic seizures and asthma attacks \cite{venegas05,mcsharry03}. Examples of sudden regime shifts 
include the collapse of vegetation under semi-arid conditions, the transition from a clear to a turbid lake, catastrophic 
shifts in fish or wildlife populations \cite{scheffer091,scheffer092,scheffer03,lenton11}, transitions from one stable 
gene expression state to another in natural and synthetic genetic circuits \cite{ozbudak04,pomerening08,ferrell01} and the 
sudden deterioration of complex diseases \cite{chen12}. The regime shifts, in most of the cases studied, are brought 
about by abrupt transitions from bistability to monostability via the saddle-node bifurcation \cite{strogatz94,scheffer091}. 
The state of a dynamical system at a specific time $t$ is defined in terms of the magnitudes of one or more key variables 
at time $t$. In the steady state, the rates of change in the magnitudes of the variables are zero. 
In the case of monostability, there is only one single stable steady state. The region of 
bistability is distinguished by the coexistence of two stable steady states separated by an unstable steady state. The two 
stable steady states correspond to low and high values of the variable, designated as the $L$ and $H$ states respectively. The 
bistable region separates two regions of monostability with the $L$ and $H$ states being the respective stable steady states. 
At a saddle node bifurcation point, one of the stable steady states merges with the unstable steady state leading to 
a reduction in the number of physical solutions, from three to one, beyond the bifurcation point.
Figure \ref{fig1} illustrates the saddle-node bifurcation in the case of the May model 
\cite{may77}, a well-known model in ecology. In the original model, a population of herbivores at a constant density 
is considered, the population being sustained by vegetation of biomass $x$. The rate equation describing the changes in 
the biomass amount is given by
\begin{eqnarray}
 \frac{dx}{dt}=rx\left(1-\frac{x}{K}\right)-\frac{cx^{2}}{x_{o}^{2}+x^{2}}
\label{model}
\end{eqnarray}
The first term on the right hand side represents the logistic growth of the 
biomass at the rate $r$, $K$ being the carrying capacity with the growth rate becoming zero when $x=K$. The second term 
corresponds to the rate of loss of the biomass due to its consumption by the herbivores. The parameter $c$ 
is the maximum loss rate of the biomass or equivalently the maximum grazing rate. The loss rate saturates when $x$ is much 
larger than a characteristic value $x_{0}$. The May model has been applied to various other ecological problems, e.g., 
the exploitation of fish populations \cite{guttal07}, dynamics of spruce budworms \cite{guttal07} and harvesting of 
macrophytes \cite{vannes02}. Figure \ref{fig1} shows the steady state values of the  
biomass $x$ $\left(\frac{dx}{dt}=0\right)$ versus 
the maximum grazing rate parameter $c$. The other parameter values are $r=1$, $K=10$ and $x_{0}=1$ in appropriate units. 
The solid lines represent the stable steady states separated by a branch (dot-dashed line) of unstable steady states. The 
parameters chosen fall in a region of parameter space exhibiting bistability. The lower $\left(c_{1}\right)$ and 
upper $\left(c_{2}\right)$ bifurcation points separate a 
region of bistability from two regions of monostability. At the upper bifurcation point $c_{2}$, 
a saddle node bifurcation takes place and there is an abrupt regime shift from the $H$ (high value of biomass) to the
$L$ (low value of biomass) state. The transition to the collapsed biomass state is not reversible with the reverse transition 
from the $L$ to the $H$ state occurring at the lower bifurcation point $c_{1}$. This is the phenomenon of hysteresis, a 
characteristic feature of regime shifts of the type shown in Figure \ref{fig1}. At $c_{1}$, a saddle node bifurcation occurs with the 
$L$ state losing stability. 

\begin{figure}
 \begin{center}
  \includegraphics[scale=0.7]{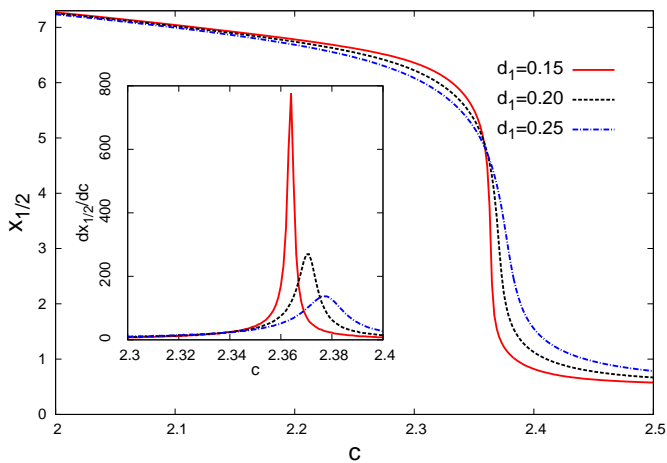}
 \end{center}
\caption{\small{The equipartition point $x_{1/2}$ (Eq. (\ref{cump3})) versus the parameter $c$ for 
three different strengths of the additive noise, $d_{1}=.15$ (solid line), 
$d_{1}=.20$ (dashed line) and $d_{1}=.25$ (dot-dashed line). The other parameter values 
are $r=1$, $K=10$ and $x_{0}=1$. The inset shows the plots of $\frac{dx_{1/2}}{dc}$ 
versus $c$ for the three noise strengths. The peak positions define the propensity 
transition point $c^{*}$.}}
\label{fig2}
\end{figure}

In the case of deterministic time evolution of a dynamical system, there are two ways in which regime shifts occur: 
\textit{(i)} at the bifurcation points and \textit{(ii)} on applying large perturbations in the region of bistability. The 
unstable steady state sets a threshold for switching transitions between the $L$ and $H$ states. Consider the system to be 
initially in the $H$ state. A sufficiently large perturbation reduces the magnitude of the dynamical variable below the 
threshold, thereby bringing about a transition to the $L$ state. 
Another way of bringing about regime shifts is through noise-induced transitions \cite{scheffer091,guttal08,dakos12,scheffer12}.
The time evolution in complex dynamical systems exhibiting 
regime shifts is, in general, stochastic in nature \cite{horsthemke84} due to the probabilistic 
nature of the processes associated with the dynamics. Stochasticity gives rise to fluctuations in the magnitude of the dynamical 
variable (say, the biomass) which, if sufficiently strong, cause transitions between the $L$ and $H$ states away from the bifurcation 
points. The occasional switch to an alternative regime in the presence of stochasticity is known as \textquotedblleft
flickering'' \cite{scheffer091} and is responsible for the appearance of a bimodal steady state probability 
distribution in the region of \textquotedblleft bistability''. Noise-induced excursions from the stable 
steady states result in a broadening of the distribution around the steady states. The steady state distribution is achieved 
when the probability of the $H$ to the $L$ state transition is the same as that of the reverse transition. The bifurcation 
theory of stochastic dynamical systems is not as well-formulated as that in the case of deterministic dynamics. Kepler 
and Elston \cite{kepler01} have defined stochastic bifurcation in terms of the number of critical (singular) points in the 
steady state probability density function. Song et. al. \cite{song10} have investigated the stochastic bifurcation structure 
of cellular networks through specifying four quantities as a function of the bifurcation parameter: \textit{(i)} the number 
of distinct subpopulations (or peaks in the steady state probability distributions), \textit{(ii)} the locations of the peaks 
in terms of the measurable variable $x$, \textit{(iii)} the variability in $x$ for each subpopulation and \textit{(iv)} the 
fractions of the whole population represented by the subpopulations. 
The total probability 
distribution is expressed as a mixture of component distributions, one for each subpopulation.     
The term subpopulation is used in a generalized sense, the number of subpopulations being equal to the number of 
distinct peaks in the probability distribution. For example, in the 
region of bistability (Figure \ref{fig1}), one has two subpopulations, $L$ and $H$ and the 
corresponding probability distributions arise due to noise-induced broadening
of the steady state levels. Guttal and Jayaprakash \cite{guttal07} have investigated
the impact of noise on bistable ecological systems. They have shown that the
region of bistability is diminished in the presence of small amounts of noise
whereas for noise beyond a critical strength, bistability vanishes altogether. 
In the latter case, the dynamical system can undergo abrupt regime shifts
frequently. 
In general, however, noise can induce new types of dynamic behaviour with significant differences 
from the deterministic dynamics \cite{horsthemke84,zakha,karma07,samoi,to10}. 
To give a few examples, Samoilov et. al. \cite{samoi} 
have shown that for the enzymatic futile cycle, a common biomolecular network motif, external noise can generate dynamic 
bistability through stochastic switchings. Zakharova et. al. \cite{zakha} have demonstrated that noise can enhance the region of 
bistability, characterized by a bimodal probability distribution between a steady state and a limit cycle, beyond the bistable
region obtained in the deterministic case. In the case of a model based on autoactivating gene expression, Karmakar and Bose \cite{karma07}
have shown that noise-induced bistability is possible in a region of parameter space for which there is no bistability 
in the deterministic case. To and Maheshri \cite{to10} have illustrated this concept in a recent experiment. 
In this paper, we propose a new framework for the quantitative
characterization of noise-induced regime shifts in bistable systems. The 
methodology has a clear physical interpretation and is easy to implement.
We focus on two models to illustrate the methodology, the May model and the genetic toggle \cite{gardner00}
describing a simple genetic circuit in which the protein products of two genes repress each other's synthesis.
The May model is a one-variable model whereas the genetic toggle is a two-variable one, thus demonstrating the 
general applicability of the methodology proposed in the paper. in Section \ref{equipartition}, we 
introduce the concept of the propensity transition point (PTP) which may be considered as an indicator of stochastic 
regime shifts. As an example, 
we consider the one-variable May model for which the steady state probability distribution is obtained by solving the 
appropriate Fokker-Planck equation. In Section \ref{gillespie}, the steady-state probability distributions are obtained 
for both the May model and the genetic toggle using the Gillespie simulation algorithm \cite{gillespie77}. We further analyze 
available experimental data to determine the PTP in the case of stochastic regime shifts from the normal 
to the persister subpopulation in mycobacteria subjected to stress in the form of nutrient depletion \cite{sureka08,ghosh11}.    
\begin{figure}
 \begin{center}
  \includegraphics[scale=0.7]{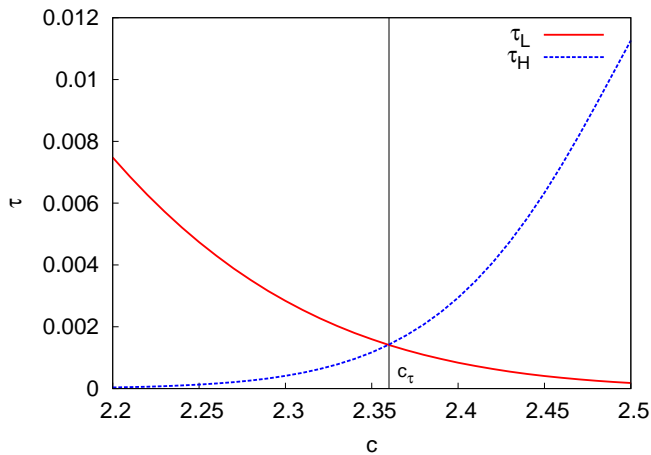}
 \end{center}
\caption{\small{The mean first passage times $\tau_{L}$ (solid line) and $\tau_{H}$ (dashed line) versus the parameter 
$c$. The other parameter values are $r=1$, $K=10$, $x_{0}=1$ and $d_{1}=.15$. The 
intersection point $c_{\tau}$ ($\tau_{L}=\tau_{H}$) of the two curves is approximately 
equal to $c^{*}$, the propensity transition point ($c_{\tau}=2.360$, $c^{*}=2.364$).}}
\label{fig3}
\end{figure}

\section{Equipartition and Propensity Transition Points}
\label{equipartition}

\begin{figure}
 \begin{center}
  \includegraphics[scale=0.7]{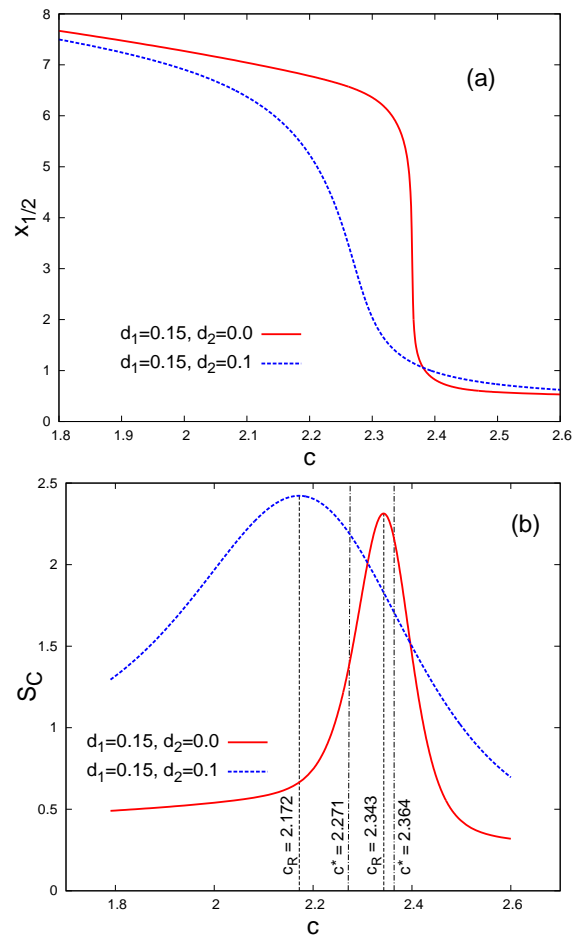}
 \end{center}
\caption{\small{(a) The equipartition point $x_{1/2}$ (Eq. (\ref{cump3})) versus $c$ for different 
strengths of the multiplicative noise: $d_{2}=0$ (solid line), $d_{2}=.1$ (dashed line) with 
$d_{1}=.15$ in each case. The other parameter values are $r=1$, $K=10$ and $x_{0}=1$.
(b) The cumulative residual entropy, $S_{C}$ (Eq. (\ref{cre})) versus $c$ for the two different 
values of the multiplicative noise strength as in (a). The PTP $c^{*}$ is included for 
comparison.}}
\label{fig4}
\end{figure}

In the case of stochastic dynamics, the probability distribution $P(x)$ of a random variable $x$ can be determined 
using a number of analytical and numerical methods \cite{gillespie77,gardiner83,kampen92}. The cumulative distribution 
function $P_{C}(x)$ is defined to be 
\begin{eqnarray}
 P_{C}(x^{\prime})=\int_{0}^{x^{\prime}}P(x)dx
\label{cump1}
\end{eqnarray}
where $0\leq x^{\prime}\leq\infty$. Let $x_{1/2}$ be the median of the distribution $P(x)$ marking 
the equipartition point of the distribution. One can write 
\begin{eqnarray}
\int_{0}^{x_{1/2}}P(x)dx=\int_{x_{1/2}}^{\infty}P(x)dx=\frac{1}{2}
\label{cump2} 
\end{eqnarray}
or,
\begin{eqnarray}
P_{C}\left(x_{1/2}\right)=1-P_{C}\left(x_{1/2}\right)=\frac{1}{2} 
\label{cump3}
\end{eqnarray}
We consider the steady-state probability distribution (SS-PD), which changes as a bifurcation parameter, e.g., the parameter 
$c$ in the May model, is changed. One can compute the variation of $x_{1/2}$ versus the bifurcation parameter 
using  Eq. (\ref{cump3}). We illustrate this in the case of May model. The SSPD, $P_{st}(x)$, 
is obtained by solving the 
Fokker-Planck equation in the steady-state. 
In the case of stochastic dynamics, the one-variable Langevin equation including both additive and multiplicative noise 
terms is given by
\begin{equation}
\frac{dx}{dt}=f(x)+g(x)\varepsilon(t)+\Gamma(t)
\label{langevin}
\end{equation}
where $\Gamma(t)$ (additive noise) and $\varepsilon(t)$ (multiplicative noise) are Gaussian white noises with zero mean and 
correlations:     
\begin{eqnarray}
\langle\Gamma(t)\Gamma(t^{\prime})\rangle&=&2d_{1}\delta(t-t^{\prime})\nonumber\\
\langle\varepsilon(t)\varepsilon(t^{\prime})\rangle&=&2d_{2}\delta(t-t^{\prime})
\label{correl}
\end{eqnarray}     
In Eq. (\ref{correl}), $d_{1}$ and $d_{2}$ denote the strengths of the additive and multiplicative noises respectively, 
we further assume that there is no correlation between the two types of noise. 
The first term in the right hand side of Eq. (\ref{langevin}) 
describes the deterministic dynamics, e.g., for the May model $f(x)$ is given by the right hand side expression in 
Eq. (\ref{model}). The Fokker-Planck equation, a rate equation for the probability distribution $P(x,t)$, is obtained from the Langevin equation as 
\cite{gardiner83,kampen92}:
\begin{equation}
\frac{\partial P(x,t)}{\partial t}=-\frac{\partial}{\partial x}[A(x)P(x,t)]+\frac{\partial^{2}}{\partial x^{2}}[B(x)P(x,t)]
\label{dist}
\end{equation}
where
\begin{equation}
A(x)=f(x)+d_{2}g(x)g^{\prime}(x)
\label{func1}
\end{equation}
and
\begin{equation}
B(x)=d_{1}+d_{2}[g(x)]^{2}
\label{func2}
\end{equation}
\noindent The SSPD is given by \cite{gardiner83,risken84,jin94} 
\begin{eqnarray}
P_{st}(x)&=&\frac{N}{B(x)}\exp\left[\int^{x}\frac{A(x)}{B(x)}dx\right]\nonumber \\
&=&\frac{N}{\{d_{2}[g(x)]^{2}+d_{1}\}^{\frac{1}{2}}}\nonumber\\
&&\times\exp\left[\int^{x}\frac{f(x^{\prime})dx^{\prime}}
{d_{2}[g(x^{\prime})]^{2}+d_{1}}\right]\nonumber\\
\;
\label{big}
\end{eqnarray}
\noindent where $N$ is the normalization constant. Equation (\ref{big}) can be rewritten in
the form
\begin{equation}
P_{st}(x)=Ne^{-\phi_{F}(x)}
\label{dist2}
\end{equation}
\noindent where
\begin{eqnarray}
\phi_{F}(x)&=&\frac{1}{2}\ln\left[d_{2}[g(x)]^{2}+d_{1}\right]\nonumber\\
&&-\int^{x}\frac{f(y)dy}{d_{2}\left[g(y)\right]^{2}+d_{1}}
\label{stochpot}
\end{eqnarray} 
defines the \textquotedblleft stochastic potential\textquotedblright$\;$ of the dynamics. The inset of Figure \ref{fig1} shows the SSPD, 
$P_{st}(x)$, versus $x$ for the May model in the presence of only additive noise ($g(x)=0$, $d_{2}=0$)
for the parameter values $r=1,\;K=10,\;x_{0}=1,\;c=2.3$ and $d_{1}=0.25$. In this case, $\phi_{F}(x)$ reduces to  
\begin{eqnarray}
 \phi_{F}(x)=\frac{1}{2}\ln d_{1}-\frac{1}{d_{1}}\int^{x}f(y)dy
\label{addpot}
\end{eqnarray}
The deterministic potential $\phi_{D}(x)$ is given by 
\begin{eqnarray}
 \phi_{D}=-\int^{x}f(y)dy
\label{detpot}
\end{eqnarray}
Comparing Eqs. (\ref{addpot}) and (\ref{detpot}) one finds that the extrema points of $\phi_{F}(x)$ and $\phi_{D}(x)$ are identical
with the minima (maximum) corresponding to stable (unstable) steady states. In the SSPD $P_{st}(x)$, the stable steady states
correspond to the maxima points $x_{1}$ and $x_{3}$ whereas the minimum point $x_{2}$ represents the unstable steady state. 
We first consider the case when only additive noise is present, i.e., $d_{1}\neq 0$ and $d_{2}=0$ in Eq. (\ref{correl}).
Figure \ref{fig2} shows the variation of $x_{1/2}$ with $c$ for the parameter values $r=1$, 
$K=10$, $x_{0}=1$ and different values of the noise strength $d_{1}=.15$ (solid line), $d_{1}=.2$ (dashed line) and 
$d_{1}=.25$ (dot-dashed line).
The deterministic bifurcation points have values $c_{1}=1.788$ and $c_{2}=2.604$. 
As $c$ is increased 
from a low value, one notices a transition of $x_{1/2}$ from a high to a low value. The transition is sharp for low values of the
additive noise and becomes smeared as the noise strength $d_{1}$ is increased. The point $x=x_{1/2}$ provides a quantitative 
measure of which subpopulation is the preferred (dominant) subpopulation in the steady state. The $L$ subpopulation corresponds to $x$ values  in the 
range $0\leq x\leq x_{2}$, the rest of the SSPD is associated with the $H$ subpopulation. For low values of $c$, $x_{1/2}$ is 
$>x_{2}$ and the $H$ subpopulation is more dominant. As $c$ increases, $x_{1/2}$ shifts towards $x_{2}$ and the $L$
subpopulation becomes the more preferred one when $x_{1/2}$ is $<x_{2}$. One can define a propensity transition point (PTP) $c^{*}$
at which the preferred subpopulation changes its character from $H$ to $L$. The inset of Figure \ref{fig2} shows the variation of 
$\frac{dx_{1/2}}{dc}$ versus $c$ for $d_{1}=.15$, $d_{1}=.2$ and $d_{1}=.25$, with the respective peak positions being 
$c^{*}=2.364$, $c^{*}=2.371$ and $c^{*}=2.378$. For larger magnitudes of noise below some critical strength, the variation 
is more smeared but one can still identify a peak position $c^{*}$ at which $\frac{dx_{1/2}}{dc}$ attains its maximum value.
In the presence of large additive noise, flickering becomes prominent and bistability is destroyed. 
When only additive noise is present and at $c=c^{*}$, $x_{1/2}$
is equal to $x_{2}$, the minimum of the SSPD. The justification of
this statement comes from the definition of $x_{1/2}$. As already
discussed, the point $x_{1/2}$ belongs to the preferred subpopulation.
At the propensity transition point $c=c^{*}$, both the subpopulations
are equally preferred, i.e., $x_{1/2}=x_{2}$, the boundary point
separating the $L$ and $H$ subpopulations. As already mentioned, in the
presence of only additive noise, the point $x_{2}$ coincides with
the unstable steady state solution in the deterministic case. 
The cumulative residual entropy defined as \cite{rao04}
\begin{eqnarray}
S_{C}=-\int_{x}P_{c}(x)\ln P_{c}(x)dx 
\label{cre}
\end{eqnarray}
with $P_{c}(x)$ as given in Eq. (\ref{cump1}), attains its maximum value at $c=c_{R}$ which is close to $c^{*}$.
For $d_{1}=.15$, $c_{R}=2.343$. Let $P_{L}$
and $P_{H}$ be the probabilities of belonging to subpopulations $L$ and $H$ respectively. $P_{L}$ and $P_{H}$ can be written 
as $P_{L}(t)=\int_{0}^{x_{2}}P(x,t)dx$ and $P_{H}(t)=\int_{x_{2}}^{\infty}P(x,t)dx$. The time evolutions of $P_{L}(t)$ and 
$P_{H}(t)$ are given by 
\begin{eqnarray}
\frac{dP_{L}(t)}{dt}&=&-k_{L}P_{L}(t)+k_{H}P_{H}(t)\nonumber \\
\frac{dP_{H}(t)}{dt}&=&-k_{H}P_{H}(t)+k_{L}P_{L}(t) 
\label{dynamics}
\end{eqnarray}
where $k_{L}$ and $k_{H}$ are the stochastic transition rates for the $L\rightarrow H$ and $H\rightarrow L$
transitions respectively. In the steady state, $\frac{dP_{L}(t)}{dt}=0$, $\frac{dP_{H}(t)}{dt}=0$  and one gets    
\begin{eqnarray}
\frac{P_{HS}}{P_{LS}}=\frac{k_{L}}{k_{H}} 
\end{eqnarray}
where \textquoteleft $S$\textquoteright$\,$ in the suffix indicates the steady state probabilities. In the case of weak noise, one can write 
\cite{zheng11}
\begin{eqnarray}
\tau_{L}&\simeq&\frac{1}{k_{L}}\nonumber\\
\tau_{H}&\simeq&\frac{1}{k_{H}}
\label{tau1} 
\end{eqnarray}
where $\tau_{L}$ and $\tau_{H}$ denotes the mean first passage times (MFPTs) for exits from the domains of the $L$ and $H$ 
subpopulations respectively. The exits are brought about by noise and a higher value of the MFPT indicates a greater 
stability of the steady state from which the exit occurs. At the PTP $c^{*}$, $P_{HS}=P_{LS}$ which leads to the equality 
$\tau_{L}=\tau_{H}$. For weak noise, $k_{H}$ and $k_{L}$ are given by the approximate expressions \cite{zheng11},
\begin{eqnarray}
k_{H}&=&\frac{d_{1}}{2\pi}\sqrt{\left|\phi^{\prime\prime}_{F}\left(x_{2}\right)\right|\phi^{\prime\prime}_{F}\left(x_{3}\right)}
e^{\phi_{F}\left(x_{3}\right)-\phi_{F}\left(x_{2}\right)}\nonumber\\
k_{L}&=&\frac{d_{1}}{2\pi}\sqrt{\left|\phi^{\prime\prime}_{F}\left(x_{2}\right)\right|\phi^{\prime\prime}_{F}\left(x_{1}\right)} 
e^{\phi_{F}\left(x_{1}\right)-\phi_{F}\left(x_{2}\right)}
\label{tau2}
\end{eqnarray}
where $\phi^{\prime\prime}_{F}\left(x_{2}\right)$ denotes the double derivative of $\phi_{F}$ with respect to $x$. Figure \ref{fig3}
shows the plots of the MFPTs $\tau_{L}$ and $\tau_{H}$ (determined using Eqs. (\ref{addpot}), (\ref{tau1}) and (\ref{tau2})) versus
the bifurcation parameter $c$. The other parameter values are $r=1$, $K=10$, $x_{0}=1$ and $d_{1}=.15$. 
The intersection point of the two curves, $c_{\tau}$, at which $\tau_{L}=\tau_{H}$, is 
approximately equal to $c^{*}$ ($c_{\tau}=2.360$, $c^{*}=2.364$). The numerical result supports the analytic argument, using Eqs. 
(\ref{dynamics}) and (\ref{tau1}), that $\tau_{L}=\tau_{H}$ at the PTP $c=c^{*}$. 

\begin{figure}
 \begin{center}
  \includegraphics[scale=0.425]{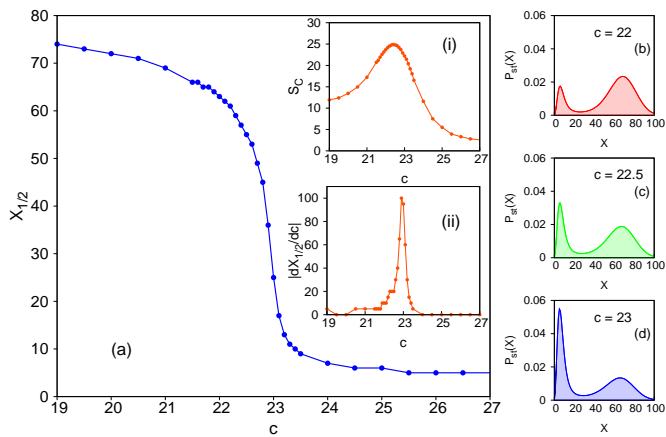}
 \end{center}
\caption{\small{(a) Variation of the equipartition point $X_{1/2}$ with the bifurcation parameter $c$. 
The graph shows a sharp fall in the values of $X_{1/2}$ in a small range of $c$ values. 
The insets (i) and (ii) show the variations 
of the cumulative residual entropy, $S_{C}$, and the first derivative of $X_{1/2}$ w.r.t. the bifurcation 
parameter $c$, $\frac{dX_{1/2}}{dc}$, respectively, as functions of $c$. $\frac{dx_{1/2}}{dc}$ 
attains a sharp maximum at $c^{*}=22.9$, the PTP. The cumulative residual entropy also becomes maximum near $c^{*}$.
The SSPDs for three values of $c$ spanning the PTP
are shown in (b),(c) and (d).}}
\label{fig5}
\end{figure}

We next consider the general case when additive and multiplicative noise terms are present in the Langevin equation, Eq. (\ref{langevin}). 
We assume that the multiplicative noise is associated with the rate constant $r$ in Eq. (\ref{model}) with $g(x)$ (Eq. 
(\ref{langevin})) given by 
\begin{eqnarray}
g(x)=x\left(1-\frac{x}{K}\right) 
\end{eqnarray}
In this case, the extrema points $x_{1}$, $x_{2}$ and $x_{3}$ of the SSPD no longer coincide with the deterministic steady state values. 
As before, the equipartition point $x_{1/2}$ can be determined as a function of $c$. 
Figure \ref{fig4} (a) shows the variation of $x_{1/2}$
versus $c$ for the parameter values $r=1$, $K=10$ and $x_{0}=1$. Figure \ref{fig4} (b) shows 
the cumulative residual entropy (Eq. (\ref{cre})) versus the bifurcation parameter $c$. The peak position $c_{R}$ is 
close to the PTP $c^{*}$ in the presence of only additive noise. 
Entropy-like quantities are measure of uncertainty 
about the values of random variables. In the case of a bimodal probability distribution, the uncertainty is maximal 
when the component subpopulations are equally preferred. In the presence of only additive noise and at $c=c^{*}$, 
the equipartition point $x_{1/2}=x_{2}$, the boundary point separating the ranges of $x$ values corresponding to 
the two subpopulations. In this case, both the subpopulations are equally preferred leading to a maximal value 
of the cumulative residual enrtopy (maximal uncertainty) close to $c^{*}$. When both additive and 
multiplicative noise terms appear in the Langevin equation (Eq. (\ref{langevin})), the difference between $c_{R}$
and $c^{*}$ increases.  The differences between the values of $x_{2}$ and $c^{*}$, the PTP, and $c_{R}$ and $c^{*}$ 
provide a measure of the multiplicative noise present in the system. 

\section{Gillespie Simulation}
\label{gillespie}

\begin{figure}
 \begin{center}
  \includegraphics[scale=0.425]{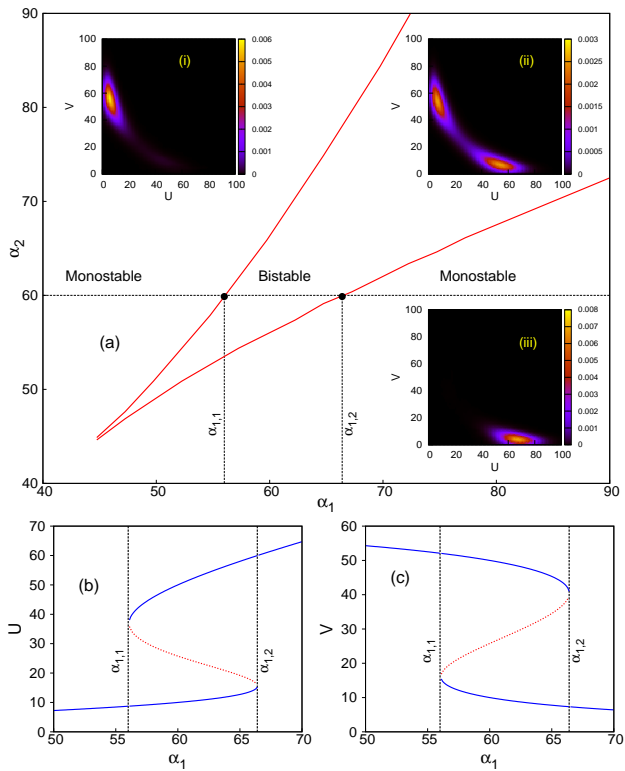}
 \end{center}
\caption{\small{(a) The $\alpha_{1}$-$\alpha_{2}$ phase diagram for the genetic toggle. The other parameter values are 
$K=500$, $\beta_{1}=\beta_{2}=2$ and $\gamma=1$. The steady-state probability distributions $P_{st}(U,V)$ for three different values of $\alpha_{1}$
are shown in the insets (i) $\alpha_{1}=50$, (ii) $\alpha_{1}=60$ and (iii) $\alpha_{1}=70$ with 
$\alpha_{2}$ kept fixed at 60. The variations of $U$ and $V$ versus the parameter $\alpha_{1}$ for $\alpha_{2}=60$ are 
shown in (b) and (c) respectively. The solid lines in each plot represent the stable steady states and the dotted line the branch 
of unstable steady states.}}
\label{fig6}
\end{figure}

The stochastic time evolution of a dynamical system can be studied
through computer simulation based on the Gillespie algorithm \cite{gillespie77}. We now discuss briefly the salient features of the Gillespie algorithm, more
detailed discussions can be obtained from Refs. \cite{gillespie77,gillespie92,scott,maini07}. 
Let $M$ be the number of reactions
controlling the time evolution of molecular numbers in a dynamical
system. Each reaction is characterized by a reaction propensity $a_{\mu}$
($\mu=1,2,\cdots,M$) with $a_{\mu}(t)dt$ defining the probability that
the reaction $\mu$ occurs in volume $V$ in the time interval ($t,t+dt$)
given the state of the system at time $t$. The propensity $a_{\mu}(t)$
is a product of two parts, the reaction rate $c_{\mu}$ for reaction
$\mu$ and the number of possible reactions $\mu$ in volume $V$.
One also defines a reaction stoichiometry matrix $\mathbf{S}$ which
is an $N\times M$ matrix where $N$ is the number of reactants (different
species of molecules) in the dynamical system.
The element $S_{ij}$ ($i=1,\cdots, N,\;\; j=1,\cdots,M$) denotes the change in the number
of reactant $i$ molecules, from $X_{i}$ to $X_{i}+S_{ij}$, when the $j$th
reaction takes place. The deterministic rate equation can be written
in a compact form
\begin{equation}
\frac{d\mathbf{X}}{dt}=\mathbf{S}.\nu
\end{equation}

The Gillespie algorithm monitors time evolution by obtaining information firstly  on the time
of occurrence of the next reaction given the state of the system in
terms of molecular numbers at time $t$ and secondly the reaction type. 
Let us define a quantity $a_{0}$ as
$a_{0}=\sum_{\mu=1}^{M}a_{\mu}$. With knowledge of the 
state of the system at time $t$, the probability that the next reaction
occurs in the time interval $t+\tau$ and $t+\tau+d\tau$ and is of
type $\mu$ is $P_{\mu}(\tau)d\tau$ where
\begin{eqnarray}
P_{\mu}(\tau)&=&a_{\mu}\exp(-a_{0}\tau)\nonumber \\
&=&\frac{a_{\mu}}{a_{0}}a_{0}\exp(-a_{0}\tau) \nonumber\\ 
&=&P_{2}(\mu)P_{1}(\tau)
\end{eqnarray}
where $P_{1}(\tau)=a_{0}\exp\left(-a_{0}\tau\right)$ and $P_{2}(\mu)=\frac{a_{\mu}}{a_{0}}$.
The Gillespie algorithm generates two random numbers $r_{1}$and $r_{2}$ using a standard
uniform random number generator. The time $\tau$ is then given by \cite{gillespie77}
\begin{eqnarray}
 \tau=\frac{1}{a_{0}}\ln\left(\frac{1}{r_{1}}\right)
\label{tau}
\end{eqnarray}
The reaction type $\mu$ is taken to be the integer for which the condition
\begin{eqnarray}
 \sum_{\nu=1}^{\mu-1}a_{\nu}<r_{2}a_{0}\leq \sum_{\nu=1}^{\mu}a_{\nu}
\label{mu}
\end{eqnarray}
is satisfied. The random number $\tau$ obtained from Eq. (\ref{tau}) is generated according to the probability 
distribution $P_{1}(\tau)$ whereas Eq. (\ref{mu}) generates the random integer $\mu$ according to the probability 
distribution $P_{2}(\mu)$. Once the pair $(\tau,\mu)$ is determined, the time $t$ is advanced by $\tau$, i.e., $t\rightarrow 
t+\tau$ and the molecular number $X_{i}$'s $(i=1,\cdots,N)$ are adjusted according to the reaction $\mu$. The SSPD $P(X_{i})$
of a molecular type can be computed by combining the data over a 
sufficiently large interval of time  after the steady state conditions are achieved. 
A rigorous approach to the study of the stochastic time evolution
of a system of $N$ chemical species participating in $M$ chemical reactions
is based on the chemical master equation \cite{gillespie77,gillespie92}. The chemical master equation 
is a time-evolution equation for the probability $P(\mathbf{x},t|\mathbf{x_{0}},t_{0})$
that the system is in the state $\mathbf{x}=(x_{1,}\cdots,x_{N})$
where $x_{i}$, $i=1,\cdots,N$, is the number of molecules of the
$i$th species at time $t$. The chemical master equation involves the reaction propensity
$a_{\mu}(\mu=1,2,\cdots,M)$ defined earlier. An exact, analytic solution
of the chemical master equation is, however, possible only in a few cases. The Gillespie
algorithm simulates the temporal trajectories of $\mathbf{x}(t)$
starting from a given initial state. The probability distribution
$P(\mathbf{x},t|\mathbf{x_{0}},t_{0})$ is computed from the knowledge
of an ensemble of sample trajectories, with the computed value approaching
the exact solution in the limit of large ensemble size. 
The \textquoteleft exact\textquoteright\, Gillespie algorithm is, however, computationally 
prohibitive when the number of chemical species, $N$, and the
number of reactions, $M$, become large.
Approximate versions of the Gillespie algorithm have been developed
\cite{gillespie92,gill13} to reduce the computational complexity. The Langevin
and Fokker-Planck equations provide approximate approaches to the
solution of the master equation and are less rigorous than the simulation
approach based on the Gillespie algorithm. Gillespie has shown \cite{gill13,gill07}
how to derive the chemical Langevin equation from the chemical master equation. The \textquotedblleft white-noise''  
form of the chemical Langevin equation is

\begin{eqnarray}
\frac{dx_{i}(t)}{dt}&=&\overset{M}{\underset{j=1}{\sum}} S_{ij}a_{j}(\mathbf{x}(t))\nonumber \\
&+&\overset{M}{\underset{j=1}{\sum}} S_{ij}\sqrt{a_{j}(\mathbf{x}(t))}N_{j}(0,1),\;
(i=1,\cdots,N)\nonumber\\
\end{eqnarray}
where $\mathbf{S}$ is the reaction stoichiometry matrix defined earlier
and $N_{j}(0,1)$ represents a unit normal random variable with mean
zero and variance one. While the chemical Langevin equation provides
an explicit structure to the noise terms in terms of the reaction
propensity, the majority of studies utilizing the Langevin formalism
start with equations of the type shown in (\ref{langevin}). The choice is dictated
by the simplicity of the calculational scheme with specific focus
on the separate effects of additive and multiplicative noise. The
Langevin equation can be demonstrated to be mathematically equivalent
to the Fokker-Planck equation \cite{gill00}. The latter is further obtained
through a Taylor series expansion of the master equation (Kramers-Moyal
expansion) and retaining terms only upto the second derivative term.

\begin{figure}
 \begin{center}
  \includegraphics[scale=0.425]{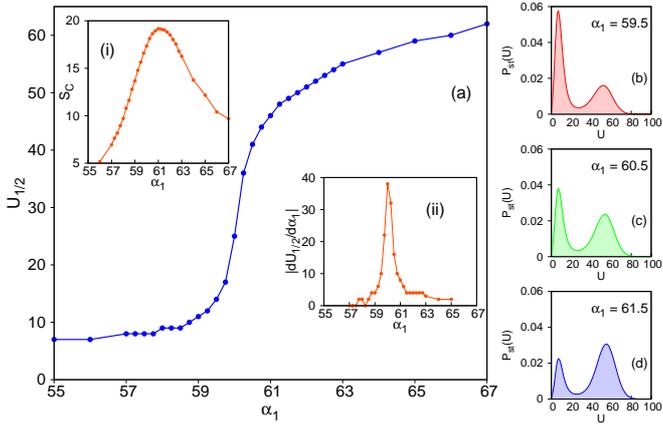}
 \end{center}
\caption{\small{(a) Variation of the equipartition point $U_{1/2}$ obtained from the SSPD
of $U$ with the bifurcation parameter $\alpha_{1}$.
The other parameter values are 
$\alpha_{2}=60$, $K=500$, $\beta_{1}=\beta_{2}=2$ and $\gamma=1$.
The graph shows a sharp fall in the values of $U_{1/2}$ in a small region of $\alpha_{1}$ values 
with decreasing $\alpha_{1}$. The insets (i) and (ii) show the variations 
of the cumulative residual entropy, $S_{C}$, and the first derivative of $U_{1/2}$ w.r.t. the bifurcation 
parameter $\alpha_{1}$, $\frac{dU_{1/2}}{d\alpha_{1}}$, respectively, as functions of $\alpha_{1}$. 
$\frac{dU_{1/2}}{d\alpha_{1}}$ 
attains a sharp maximum at $\alpha_{1}^{*}=60$, the PTP. The cumulative residual entropy also becomes maximum near $\alpha_{1}^{*}$.
The SSPDs for three values of $\alpha_{1}$ spanning the PTP are
shown in (b),(c) and (d).}}
\label{fig7}
\end{figure}

We next report the results of the Gillespie simulation in the cases
of the May model and the genetic toggle. The simulation tracks stochastic time evolution without 
distinguishing between additive and multiplicative types of noise.
The rigorous Gillespie simulation approach is valid only for unimolecular
and bimolecular reactions \cite{gill13}. Since the simulation approach
is computationally exhaustive, approximation versions of the algorithm
are often in use. One such approximation involves the study of a reduced
model for which each reaction is a composite one representing the
combined effects of multiple elementary reactions \cite{scott}. The reaction
propensity $a_{\mu}(\mu=1,2,\cdots,M)$ and the $N\times M$ stoichiometry
matrix $\mathbf{S}$ are defined for a system of $M$ composite reactions
in which $N$ chemical species participate. The approximate approach,
though less rigorous, reproduces qualitatively the results obtained
using the rigorous simulation approach \cite{scott}.
For the May model, we have
two composite reactions each of which effectively represents a number
of more elementary reactions. Rewriting the biomass molecular number
as $X$, the rate equation (\ref{model}) can be recast as
\begin{equation}
\frac{dx}{dt}=F_{1}(X)-F_{2}(X)
\end{equation}
with
\begin{eqnarray}
F_{1}(X)&=&rX\nonumber \\
F_{2}(X)&=&\frac{rX^{2}}{K}+\frac{cX^{2}}{X_{0}^{2}+X^{2}}
\end{eqnarray}
There are thus two composite reactions $X\longrightarrow X+1$ and
$X\longrightarrow X-1$ with the reaction propensities given by $F_{1}(X)$
and $F_{2}(X)$ respectively. The stoichiometry matrix $\mathbf{S}$
is of size $(1\times2)$ with $S_{11}=+1$ and $S_{12}=-1$. Figure
\ref{fig5}(a) shows the variation of the equipartition point $X_{1/2}$ with
the bifurcation parameter $c$. The other parameter values are $r=1$,
$K=100$ and $X_{0}=10$. One notes the sharp fall in the values of $X_{1/2}$
in a small region of $c$ values. The required SSPD's are computed
by collecting together the data in the Gillespie simulation over a large period of time.
The insets \textit{(i)} and \textit{(ii)} of Figure \ref{fig5}(a) show the variations
of the cumulative residual entropy, $S_{c}$, and the first derivative $\frac{dX_{1/2}}{dc}$
respectively as functions of the bifurcation parameter $c$. The first
derivative attains a sharp maximum at $c^{*}=22.9$, the PTP. The
cumulative residual entropy, $S_{c}$, has its maximum value close to $c^{*}$. Figures \ref{fig5}(b), (c) and
(d) show the SSPDs for three values of the bifurcation parameter spanning
the PTP.

We next describe the results for the genetic toggle, a two-variable
model. The toggle circuit consists of two genes synthesizing proteins
with molecular numbers $U$ and $V$. The proteins repress each others'
synthesis. The dynamics of the model are described by the set of equations \cite{gardner00}
\begin{eqnarray}
\frac{dU}{dt}&=&\frac{\alpha_{1}}{1+\frac{V^{\beta_{1}}}{K}}-\gamma U \nonumber \\
&=&F_{1}(U,V)-F_{2}(U,V)
\end{eqnarray}
\begin{eqnarray}
\frac{dV}{dt}&=&\frac{\alpha_{2}}{1+\frac{U^{\beta_{2}}}{K}}-\gamma V \nonumber\\
&=&G_{1}(U,V)-G_{2}(U,V)
\end{eqnarray}
The parameters $\alpha_{1},\alpha_{2}$ denote the effective rates
of synthesis and $\gamma$ the degradation rate constant, assumed
to be the same for the two proteins. The parameters $\beta_{1}$ and
$\beta_{2}$ are the indices indicating cooperativity in repression
and $K$ is related to the binding constant of proteins. The functions
$F_{i}(U,V)$ and $G_{i}(U,V)$ ($i=1,2$) are:
\begin{eqnarray}
F_{1}(U,V)&=&\frac{\alpha_{2}}{1+\frac{U^{\beta_{2}}}{K}},\;F_{2}(U,V)=\gamma U \nonumber \\
G_{1}(U,V)&=&\frac{\alpha_{1}}{1+\frac{V^{\beta_{1}}}{K}},\;G_{2}(U,V)=\gamma V
\end{eqnarray}
There are now four composite reactions ($M=4$) with reaction propensities
$\mu_{i}$, $i=1,\cdots,4$. The stoichiometry matrix $\mathbf{S}$ is
of size $2\times4$. The composite reaction scheme is:
\begin{eqnarray}
U&\longrightarrow& U+1,\;\mu_{1}=F_{1}(U,V)\nonumber \\
U&\longrightarrow& U-1,\;\mu_{2}=F_{2}(U,V)\nonumber \\
V&\longrightarrow& V+1,\;\mu_{3}=G_{1}(U,V)\nonumber \\
V&\longrightarrow& V-1,\;\mu_{4}=G_{2}(U,V)
\end{eqnarray}
Also,
\begin{equation}
\mathbf{S}=\left(\begin{array}{cccc}
1 & -1 & 0 & 0\\
0 & 0 & 1 & -1\\
\end{array}\right)
\end{equation}
Figure \ref{fig6} shows the $\alpha_{1}$-$\alpha_{2}$ phase diagram of the genetic
toggle exhibiting two regions of monostability and one region of bistability.
The other parameter values are: $K=500$, $\beta_{1}=\beta_{2}=2$
and $\gamma=1$. The SSPDs $P_{st}(U,V)$ for three different values
of $\alpha_{1}$ and with $\alpha_{2}=60$ are shown in the insets
\textit{(i)} $\alpha_{1}=50$, \textit{(ii)} $\alpha_{1}=60$ and \textit{(iii)}
$\alpha_{1}=70$. These have been computed using the Gillespie simulation
algorithm. The variations of the steady state values of $U$ and $V$
versus the parameter $\alpha_{1}$ ($\alpha_{2}=60$) are shown in
Figures \ref{fig6}(b) and \ref{fig6}(c) respectively. The branches of stable steady
states (solid lines) are separated by the branches of unstable steady
states (dotted lines). Figure \ref{fig7} exhibits the results obtained from the Gillespie
simulation. Figure \ref{fig7}(a) shows the variation of the equipartition point
$U_{1/2}$, obtained from the SSPD of $U$, versus the bifurcation
parameter $\alpha_{1}$. The other parameter values are $\alpha_{2}=60$,
$K=500$, $\beta_{1}=\beta_{2}=2$ and $\gamma=1$. The SSPDs are
obtained by collecting data over $10^{8}$ time points in the steady
state. One notes a sharp fall in the values of $U_{1/2}$ in a small
region of $\alpha_{1}$ values when $\alpha_{1}$ is decreased from high to
low values. The insets \textit{(i)} and \textit{(ii)} exhibit the variations
of the cumulative residual entropy, $S_{c}$, and $\frac{dU_{1/2}}{d\alpha_{1}}$ respectively
versus the bifurcation parameter $\alpha_{1}$. $\frac{dU_{1/2}}{d\alpha_{1}}$
attains a sharp maximum at $\alpha_{1}^{*}=60$, the PTP. The cumulative residual entropy
also attains its maximum values for $\alpha_{1}$ close to $\alpha_{1}^{*}$.
The SSPDs for three values of $\alpha_{1}$, spanning the PTP, are
shown in Figures \ref{fig7}(b), (c) and (d).

\begin{figure}
 \begin{center}
  \includegraphics[scale=0.65]{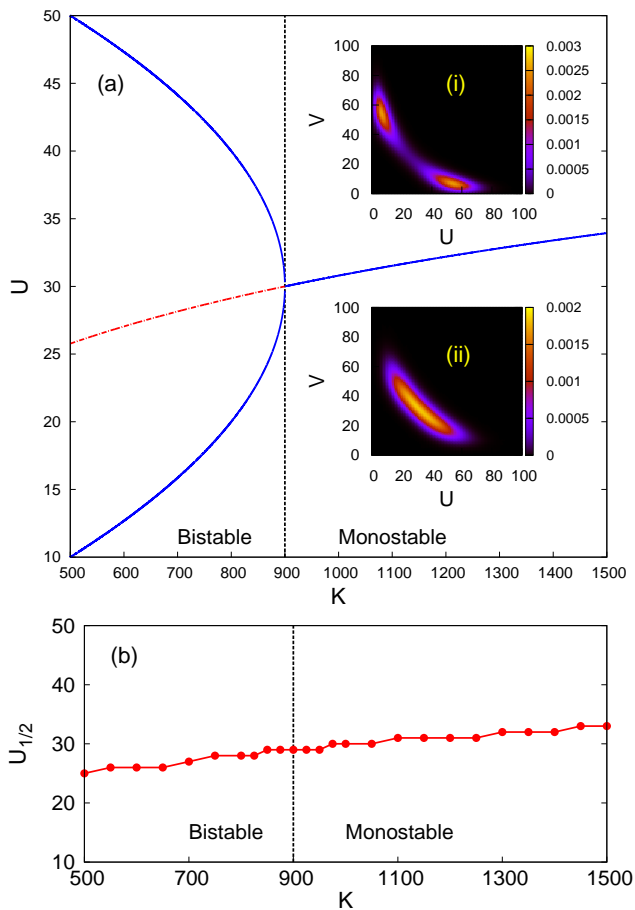}
 \end{center}
\caption{\small{(a) The genetic toggle model exihibits pitchfork bifurcation with respect to 
the parameter $K$ when $\alpha_{1}=\alpha_{2}$. The parameter values are 
$\alpha_{1}=\alpha_{2}=60$, $\beta_{1}=\beta_{2}=2$ and $\gamma=1$. 
The pitchfork bifurcation point is $K=900$. The model shows bistability for 
$K<900$ and monostability for $K>900$. The insets show SSPDs for the cases (i) $K=500$ 
and (ii) $K=1200$. (b) The variation of $U_{1/2}$ against the bifurcation 
parameter $K$. The plot does not provide a signature of stochastic bifurcation. }}
\label{fig8}
\end{figure}

The sharp fall of the equipartition point (median) $U_{1/2}$ in the vicinity of the PTP is a unique characteristic of the 
saddle-node bifurcation. The pitchfork bifurcation is another type of bifurcation involving a transition between 
bistability and monostability. The genetic toggle exhibits pitchfork bifurcation as a function of the parameter $K$ 
when the effective synthesis rate constants $\alpha_{1}$ and $\alpha_{2}$ are kept equal. Figure \ref{fig8}(a) 
demonstrates the bifurcation with the steady state value of $U$ plotted against $K$. The solid lines represent stable 
steady state and the dot-dashed line the unstable steady state. The parameter values are $\alpha_{1}=\alpha_{2}=60$, 
$\beta_{1},\beta_{2}=2$ and $\gamma=1$. The pitchfork bifurcation point $K=900$ separates a region of bistability 
$(K<900)$ from a region of monostability $(K>900)$. The insets of Figure \ref{fig8}(a) show the SSPDs determined 
using the Gillespie algorithm for \textit{(i)} $K=500$ and \textit{(ii)} $K=1200$. Figure \ref{fig8}(b) shows the variation of $U_{1/2}$
against $K$. In this case, there is no distinguishing feature in the plot providing a signature of stochastic 
bifurcation. The numerical evidence of a propensity transition thus enables one to distinguish
between the pitchfork and saddle node bifurcations as the basis for noise-induced
regime shifts. We note here that since the propensity transition point as an indicator
of stochastic regime shifts is not an appropriate indicator of the
pitchfork bifurcation, it cannot also distinguish between the cases
of pitchfork bifurcation and the absence of bifurcation. In the latter
case, since only one population is present, the variation of the equipartition
point as a function of the bifurcation parameter is expected to be
featureless, a trait shared with the pitchfork bifurcation.

There is recent experimental evidence that microorganisms take recourse
to noise-induced regime shifts as a strategy for survival under stress
\cite{sureka08,ghosh11,smits06,maamar07}. In the case of mycobacteria, stresses like nutrient depletion
activate the stringent response pathway involving the two component
system MprAB, the sigma factor SigE and the stringent response regulator
Rel. The enzyme polyphosphate kinase 1 (PPK1) regulates the pathway
by catalysing the synthesis of polyphosphate required for the activation
of MprB. Multiple positive feedback loops and molecular sequestration
of the sigma factor \cite{sureka08,tiwari10} create the potential for bistability,
i.e., alternative attractor states. Stochastic gene expression gives
rise to fluctuations in the levels of key regulatory proteins like
Rel which, if sufficiently strong, can bring about regime shifts from
one attractor state (low Rel level) to another (high Rel level). The
total mycobacterial population divides into two distinct subpopulations
$L$ and $H$ corresponding to low and high Rel levels respectively. The
stringent response pathway, activated by Rel proteins, is initiated
in the $H$ subpopulation. It is this subpopulation which adapts to stress
and is designated as the persister subpopulation. The other subpopulation,
termed the normal subpopulation, is unable to survive under stress.
When exposed to antibiotic drugs, the normal subpopulation is killed
while the persisters are able to survive. The persisters have slow
growth rates and low rates of metabolic activity. They stay dormant
for long periods of time waiting for the opportune moment to revive
and restart the mycobacterial infection. Single cell analysis via
flow cytometry provides experimental evidence of regime shifts from
the $L$ to the $H$ subpopulation in a mycobacterial population subjected
to nutrient depletion till the stationary phase is reached \cite{sureka08,ghosh11}.
The experiments were carried out on \textit{M.smegmatis} which shares
similar genetic circuitry with \textit{M.tuberculosis}, the pathogen
responsible for tubercular infection. The existence of persisters
acts against the total eradication of the tubercular infection. The
major goal of an effective drug treatment is to eliminate the population
of persisters. In this context, it is of interest to identify the
PTP above which the subpopulation of persisters becomes the dominant
subpopulation.

\begin{figure}
\begin{center}
 \includegraphics[scale=0.425]{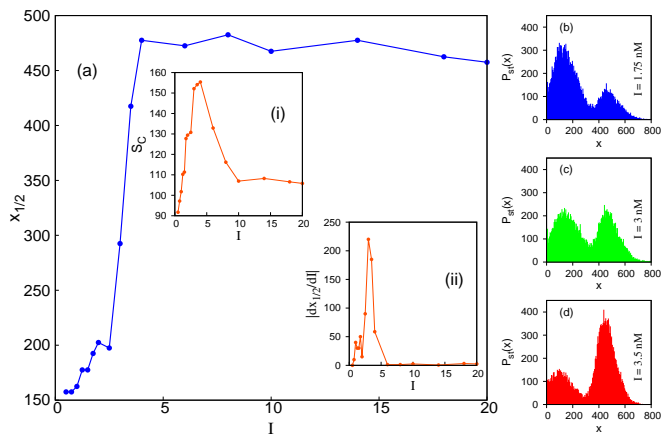}
\end{center}
\caption{\small{(a) Variation of equipartition point $x_{1/2}$ 
versus inducer concentration $I$ (in nM). The insets show the variation
 of (i) the cumulative residual entropy, $S_{C}$, and (ii) $\left|\frac{dx_{1/2}}{dI}\right|$ versus the inducer concentration
 $I$. (b)$-$(d) Stationary phase probability distributions of Rel levels
 for $I=1.75$ nM, 3 nM and 3.5 nM respectively.}}
\label{fig9}
\end{figure}

The stochastic bifurcation analysis is carried out on available experimental
data \cite{sureka08,sureka} with the \textit{ppk1} gene designed to be tetracycline-inducible.
Figure \ref{fig9}(b), (c) and (d) shows the distribution of Rel levels in the stationary
phase of a population of \textit{M.smegmatis} cells for three different
inducer concentrations $I$ (in nM). Figure \ref{fig9}(a) shows the variation
of the equipartition point $x_{1/2}$ as a function of $I$ (in nM),
the inducer concentration. The insets \textit{(i)} and \textit{(ii)} show the variations
of $\left|\frac{dx_{1/2}}{dI}\right|$ and the cumulative residual entropy, $S_{c}$, respectively versus
the inducer concentration $I$. The propensity transition to the persister
subpopulation is found to occur at low values of $I$.

\section{Concluding Remarks}
\label{colclude}
The quantitative measure of regime shifts, namely, the PTP, proposed in this Letter is of broad applicability 
and can be computed both when the model dynamics are known or when only the time series data
are available. The physical interpretation of the PTP in terms of a tipping of the balance 
towards one regime or the other helps in identifying the parameter regime to be avoided in order to forestall 
disastrous regime shifts. The variation of the equipartition point $x_{1/2}$ versus a bifurcation parameter provides 
advanced knowledge of the approach to the PTP. A number of signatures of impending regime shifts have been proposed so 
far in the scenario 
of the saddle-node bifurcation \cite{strogatz94,scheffer091,guttal08,dakos12}. These include the critical slowing down,
rising variance and skewness of a subpopulation 
probability distribution and the rising lag-$1$ auto-correlation function, as a function of the bifurcation parameter. 
If the model governing the dynamics of the system of interest is known, one can calculate the return time (the time 
taken by a system to regain its stable steady state after the system is weakly perturbed) using the standard procedure 
of linear stability analysis \cite{strogatz94,wissel84}. The return time diverges (critical slowing down) as a bifurcation point at which the 
steady state loses its stability is approached. A variety of techniques, e.g., the linear noise approximation \cite{kampen92}
can be used to calculate quantities like the variance and the lag-$1$ autocorrelation function. The demonstration of the 
skewness of a subpopulation probability distribution as the bifurcation point is approached requires the use of somewhat 
ad-hoc procedures for obtaining a component probability distribution from the total distribution \cite{guttal07,guttal08}. One of these is 
to introduce suitable cut-off procedures to isolate a component distribution which exhibits rising skewness as a 
bifurcation point is approached.  
The formalism developed in this paper for a quantitative characterization of noise-induced regime shifts is applicable even when 
only time series data are available, without an adequate knowledge of the underlying model. To make the analysis possible, 
one has to identify a parameter $\theta$ which is a relevant bifurcation parameter. If time series data are available over a 
sufficiently long interval of time, for which $\theta$ remains constant, one can construct the probability distribution of the 
relevant variable from the data itself. Similar probability distributions can be computed for other values of $\theta$ if it is 
a changing function of time. Several studies have recently been undertaken on the early signatures of sudden regime shifts
\cite{scheffer091,scheffer092,scheffer12} when only time series data are available.
As already discussed in the Introduction, regime shifts are 
often noise-induced. Such transitions, depending on the magnitude of the noise, may occur even when the system is not close to 
a bifurcation point. The frequency of the transitions is expected to increase in the vicinity of the bifurcation point. The 
method developed in the present study is based on the concept of the PTP which precedes the bifurcation point. The sharp fall 
of the equipartition point $x_{1/2}$ in the vicinity of the PTP also serves as an indicator of the system approaching a 
bifurcation point apart from signifying a stochastic shift in the propensity of the system to be in one of two alternative 
regimes. The utility of the PTP formalism has been demonstrated for both one-variable and two-variable models. The $x_{1/2}$
versus bifurcation parameter plot further carries a distinctive signature of the saddle-node bifurcation when the plot exhibits 
a sharp fall. The sharp fall of $x_{1/2}$ in the vicinity of the PTP is an indicator of a stochastic regime shift. Regime 
shifts are of common occurrence in wide ranging dynamical systems. The shifts may be due to changing parameters resulting in 
a bifurcation or due to a large perturbation causing a switch between one attractor state to another across the border 
separating the basins of attraction of the attractors or due to noisy dynamics with the fluctuations in key variables driving 
the regime shift. A large number of studies \cite{scheffer091,scheffer092,scheffer12} provide a quantitative characterization 
of sudden regime shifts at the bifurcation points. The present study focuses on a quantitative indicator of stochastic regime 
shifts which has a simple physical interpretation and is straightforward to compute.

\subsection*{Acknowledgements}

SG acknowledges the support of CSIR, India, under Grant No. 09/015(0361)/2009-EMR-I.

\singlespacing

\end{document}